%% file: FinnTro.tex
\documentclass[conference]{IEEEtran}
\usepackage{amsmath,amssymb,amsfonts}
\usepackage{multirow}

\usepackage{algorithm}
\usepackage{array}
\usepackage[caption=false,font=normalsize,labelfont=sf,textfont=sf]{subfig}
\usepackage{textcomp}
\usepackage{stfloats}
\usepackage{url}
\usepackage{verbatim}
\usepackage{graphicx}
\usepackage{booktabs}
\usepackage{cite}
\usepackage{xcolor}
\usepackage[nice]{nicefrac}
\usepackage[colorlinks=blue,bookmarksopen,bookmarksnumbered,citecolor=blue,urlcolor=blue]{hyperref}
\usepackage{algpseudocode}
\usepackage{amsmath}
\usepackage{amssymb}
\usepackage{algorithm}
\usepackage{float}
\usepackage{algorithmicx}

\begin{document}
\newcommand{\BIBentryALTinterwordspacing}{%
  \spaceskip=\fontdimen2\font plus \fontdimen3\font minus \fontdimen4\font\relax}
\newcommand{\BIBentrySTDinterwordspacing}
{\spaceskip=0pt\relax}

\title{FINN-Tro: Exploiting Verification Gaps in Dataflow Inference Accelerators}

\author{Qazi Arbab Ahmed, Suraj Karki, Thorsten Jungeblut\\
\IEEEauthorblockA{Bielefeld University of Applied Sciences and Arts (HSBI), Bielefeld, Germany
}
Email: \{qazi.ahmed, suraj.karki, thorsten.jungeblut\}@hsbi.de
}




\maketitle

\begin{abstract}
The growing adoption of dataflow accelerators for neural network inference introduces new attack surfaces that existing verification methodologies fail to address. Inference acceleration frameworks such as FINN, which transform quantized neural networks into FPGA-deployable dataflow architectures, implicitly assume semantic equivalence between the software model and the synthesized hardware. In this work, we introduce the FINN-Tro attack, which identifies and exploits a critical verification gap in the FINN compilation pipeline that enables stealthy hardware Trojan insertion without modifying the original quantized model. The Trojan is placed in the last Matrix-Vector Activation Unit (MVAU) layer and supports two counter-based trigger modes, periodic and persistent, and three payload types: bias addition, logit swapping, and bias subtraction, resulting in six different configurations. FINN-Tro is evaluated on an MNIST feed-forward network and a CIFAR-10 convolutional neural network deployed on a PYNQ-Z1 board. Across the evaluated configurations, accuracy reductions range from 0.90\% to 82.84\%, while throughput and runtime remain close to the corresponding baseline designs. The most severe configuration, persistent \textit{Bias Addition}, reduces accuracy from 92.96\% to 10.12\% on MNIST and from 84.19\% to 10.00\% on CIFAR-10. The inserted logic introduces modest implementation overhead, with maximum LUT and FF increases of 6.71\% and 7.49\% for MNIST, and 2.50\% and 3.98\% for CIFAR-10, respectively. Our findings reveal that widely used pre- and post-compilation verification flows are insufficient for detecting such temporally delayed hardware manipulations, motivating the need for stronger verification mechanisms in accelerator toolchains.


\end{abstract}


\section{Introduction}
\label{sec:Intro}

\input{1_Introduction}
\section{Related Work}
\label{sec:literature}
\input{2_Related_Work}

\section{Methodology}
\label{sec:PropMeth}
\input{3_Proposed_Scheme}

\section{Results}
\label{sec:results}
\input{4_Results}

\section{Conclusion}
\label{sec:Con}
\input{5_Conclusion}

\section*{Acknowledgments}
\small
We gratefully acknowledge funding by the projects enableATO (German Federal Ministry of Transport, grant: 19DZ23002D), FH-Personal (German Federal Ministry of Research, Technology and Space (BMFTR), grant: 03FHP106) and KI-Akademie OWL (BMFTR, supported by VDI/VDE Innovation+Technik GmbH, grant: 16IS24057C) and appreciate the discussions with the EKI project (https://www.eki-project.tech/) members at Paderborn University.

\bibliographystyle{IEEEtran} 


\end{document}

%% file: 1_Introduction.tex
Deep neural networks (DNNs) are increasingly used in safety-critical applications, including industrial inspection, medical diagnostics, and autonomous driving, where fast and reliable inference is essential but computationally expensive. Field-Programmable Gate Arrays (FPGAs) provide a platform for DNN acceleration by enabling highly parallel and reconfigurable hardware architectures with lower latency and greater flexibility than general-purpose processors.

FPGA-based DNN accelerators primarily use systolic and dataflow architectures. Systolic architectures share processing units across layers over time, whereas dataflow architectures assign dedicated processing units to individual layers, enabling fine-grained parallelism and streaming execution. Open-source third-party compiler frameworks such as FINN \cite{10.1145/3020078.3021744} and hls4ml \cite{10.1145/3801979} simplify FPGA deployment by automatically generating hardware accelerators from trained neural networks. However, reliance on third-party compilation toolchains introduces security and trust concerns, particularly when the generated hardware is not rigorously verified against the original software model \cite{lomet:hal-05120223}, \cite{11215781}.
\begin{figure}[!htbp]
    \centering
    \includegraphics[width=\columnwidth]{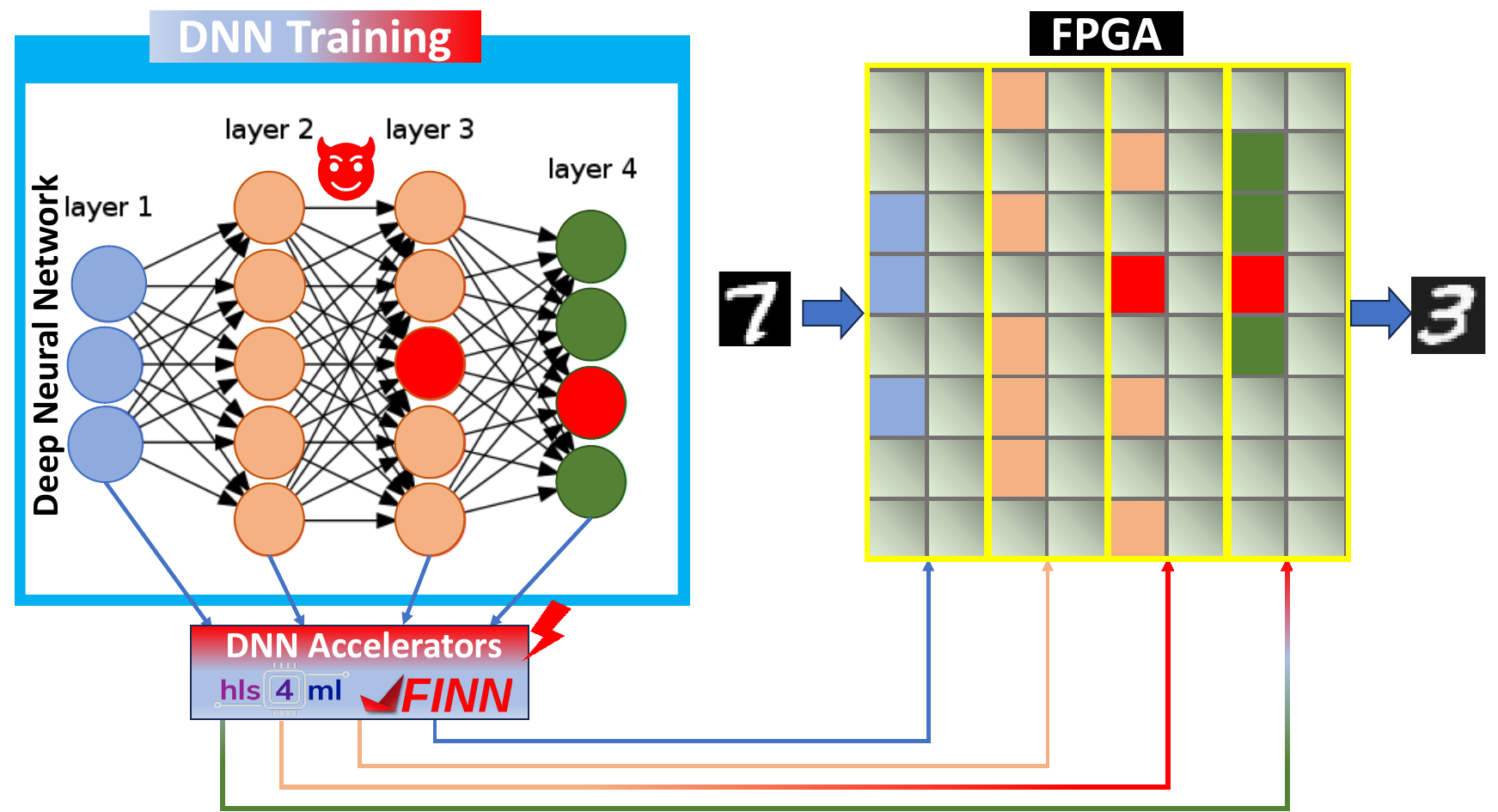}
    \caption{Security threats to FPGA-based DNN accelerators causing hidden-layer manipulation and misclassification during inference.}
    \label{fig:threatmodel}
\end{figure}
Figure \ref{fig:threatmodel} illustrates a hardware Trojan (HT) attack in which malicious logic is introduced into a third-party FPGA-based DNN accelerator to manipulate intermediate computations and cause misclassification. Such an attack can compromise accelerator integrity without modifying the trained DNN model. In this work, we use an adapted version of the FINN compiler as the target DNN accelerator-generation framework.


FINN validates generated accelerators primarily through functional simulation, comparing an output against a golden reference at selected build steps. However, this validation approach is structurally incapable of detecting malicious logic that is triggered by temporal conditions and specific inputs. A rogue employee in the development team, or a supply-chain attacker who controls any component of the FINN compilation environment, can exploit this verification gap by inserting hardware Trojan logic into the generated HLS C++ code.

To investigate this vulnerability, we propose FINN-Tro, a systematic methodology for inserting hardware Trojans into FINN-generated FPGA accelerators. To the best of our knowledge, FINN-Tro is the first demonstrated hardware Trojan attack targeting the FINN compilation pipeline. The main contributions are:

\begin{itemize}
    \item The proposed Trojan is embedded in the final Matrix-Vector Activation Unit (MVAU), which generates the classification output logits. It employs a counter-based trigger with two modes, Periodic and Persistent, and three configurable payloads: Bias Addition, Logit Swap, and Bias Subtraction.

   \item FINN-Tro is evaluated across six attack configurations using two DNN architectures: a feed-forward network for MNIST digit recognition and a convolutional neural network (CNN) for CIFAR-10 classification, both deployed on a Xilinx PYNQ-Z1 FPGA.

   \item  Our results show that, depending on the configuration, FINN-Tro causes accuracy degradation ranging from 0.90\% to 82.84\%, while incurring only a small overhead and remaining undetectable by FINN functional simulation.
\end{itemize}

We, therefore, identify and exploit a critical verification gap in the FINN compilation pipeline, demonstrating that the existing compilation flow lacks the necessary mechanisms to detect malicious hardware-level modifications introduced during layer specialization.

%% file: 2_Related_Work.tex
\begin{figure*}[bp]
    \centering
    \includegraphics[width=\linewidth]{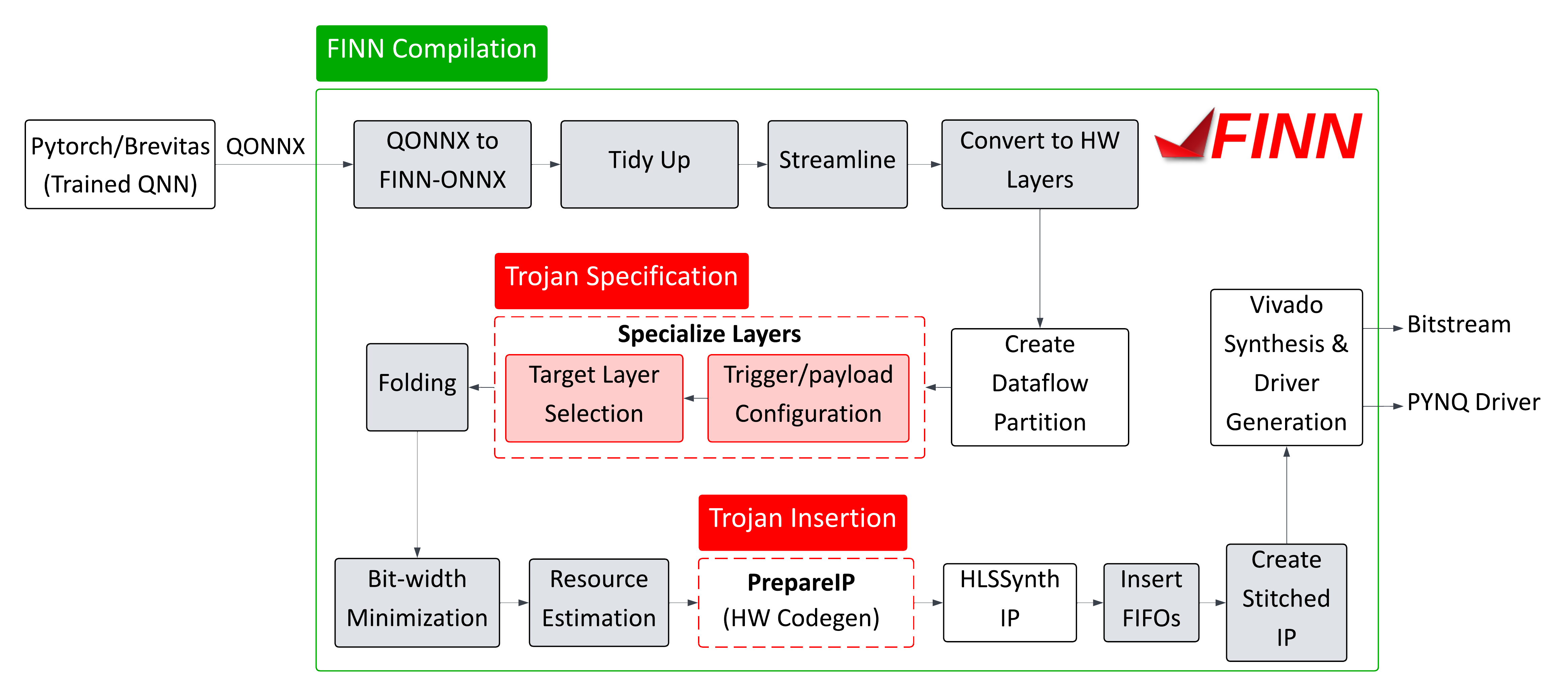}
    \caption{Overview of FINN-Tro, the proposed hardware Trojan insertion method built on the FINN compilation flow}
    \label{fig:trojan}
\end{figure*}
To achieve high-throughput and low-latency DNN inference, Field-Programmable Gate Arrays (FPGAs) are increasingly used for AI applications \cite{10918097}. Compilers such as FINN, Vitis AI, and Gemmini automate mapping quantized DNNs onto FPGA-based dataflow or systolic accelerators. Two common architectures are sequential and fine-grained. Sequential accelerators, such as the AMD Xilinx Vitis AI DPU, process layers one at a time, providing flexibility and hardware reuse but limiting throughput \cite{jlpea12020030}. In contrast, fine-grained architectures such as FINN \cite{10.1145/3020078.3021744} and hls4ml \cite{10.1145/3801979} exploit layer-level parallelism and streaming, achieving state-of-the-art performance for real-time applications \cite{10017269}. However, third-party automated hardware-generation frameworks typically lack robust equivalence verification between the original software model and generated HLS/RTL implementations \cite{pch_ahmed}, creating opportunities for hardware-level attacks. Lomet et al. \cite{lomet:hal-05120223} demonstrated passive extraction of FINN accelerator parameters, recovering folding factors and quantization levels with over 95\% accuracy from on-chip power traces. FPGAs are also susceptible to fault injection and bit-flip attacks; resilience analysis showed that faults in internal weights, particularly MSBs, can severely degrade model accuracy \cite{pollo2023resilience}.

Hardware Trojans (HTs) pose a further threat by modifying accelerator hardware while leaving the DNN model unchanged \cite{dong2020hardware}. Prior work has shown Trojan attacks at different steps of FPGA-based AI accelerator deployment, including the design flow, bitstream, and post-configuration steps \cite{mlut, Ahmed_DATE, ahmed2024post}. Warnecke et al. \cite{10918097} demonstrated a Trojan in the LOAD engine of a Xilinx Vitis AI DPU that intercepts selected model parameters, causing misclassification with only 0.24\% circuit overhead. Shahzad et al. \cite{11228721} introduced ShrinkBox, which manipulates object-detection bounding boxes under a specific trigger. This causes significant distance-estimation errors in autonomous driving systems, leading systems to fail or delay warnings. Other recent attacks target internal accelerator structures, including Transformer \cite{11495516} and SNN \cite{11050044} accelerators, achieving severe accuracy degradation with very low hardware overhead. Bhuekar \cite{Bhuekar2025} further demonstrated cross-layer attacks in automated accelerator-generation frameworks such as Gemmini, using hidden instructions to trigger bit-level parameter modifications.

However, existing approaches do not consider the specific verification gap introduced by third-party custom dataflow accelerators such as FINN, where malicious logic can be inserted without modifying the original DNN model. We propose a novel hardware Trojan attack, called FINN-Tro, which is designed for the MVAU layer of the FINN flow. It uses a counter-based trigger with six different configurations. The following section presents a detailed methodology of our FINN-Tro attack.

%% file: 3_Proposed_Scheme.tex
Our proposed methodology targets the FINN compilation pipeline used to generate FPGA-based DNN accelerators, as illustrated in Fig. \ref{fig:trojan}. The attacker has access to the HLS code-generation stage, where malicious logic is inserted into the synthesizable HLS C++ code generated for an MVAU layer. The proposed method consists of two phases: i) Trojan specification and ii) Trojan insertion.
\begin{figure*}[ht]
    \centering
    \includegraphics[width=\linewidth]{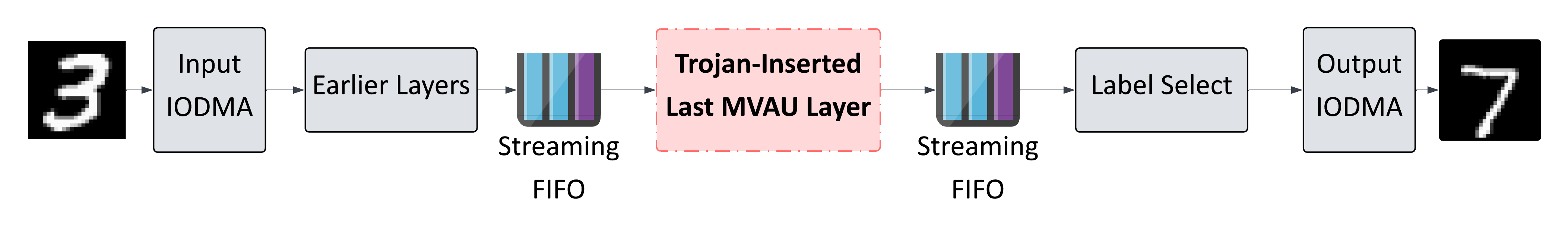}
    \caption{Placement of the hardware Trojan within the FINN-generated accelerator’s inference pipeline}
    \label{fig:trojan_placement}
\end{figure*}
\subsection{Threat Model}
We assume an adversary capable of modifying the FINN compilation flow and its software components, particularly the layer-specialization module and HLS code-generation backend, without access to the user's training dataset or trained model. The objective is to cause controlled inference misbehavior under specified trigger conditions while preserving normal behavior otherwise. This attack model exploits the gap between model-level functional verification and the final hardware implementation.

\subsection{Trojan Insertion in FINN Data-flow Build}
FINN provides an end-to-end compilation flow for implementing quantized neural networks on FPGAs \cite{10.1145/3020078.3021744}. The input quantized model is converted into the ``FINN-ONNX'' intermediate representation, which is subsequently transformed through graph-level optimizations and hardware-specific transformations. Functional verification can be performed during these transformations to check whether model behavior is preserved. However, after the graph is lowered to hardware-specific layers and HLS code, the representation becomes tightly coupled to the hardware implementation.

FINN-Tro exploits this stage by inserting Trojan information during layer specialization and realizing the corresponding malicious logic during HLS code generation. Specifically, Trojan trigger and payload attributes are attached to the target MVAU node during layer specialization. The PrepareIP backend then reads these attributes when generating the HLS C++ template and incorporates the Trojan logic into the resulting MVAU IP. The quantized model graph and trained weights remain unchanged; the modification is limited to the HLS generation stage. Consequently, FINN's verification supports only a single input and resets the design state for each verification, preventing detection of the inserted malicious behavior at any pipeline stage. Algorithm~\ref{alg:trojan_insertion} summarizes the insertion process.

\begin{algorithm}[!htbp]
\caption{: Trojan insertion during HLS code generation}
\label{alg:trojan_insertion}
\begin{algorithmic}[1]

\Require ONNX model $G$, FPGA target $\theta$, Trojan configuration $C$
\Ensure Bitstream $B$ with embedded hardware Trojan

\State $G_{HW} \gets \text{FINN\_preprocessing}(G, \theta)$
\State $M \gets \text{select\_trojan\_mvau}(G_{HW}, C.\text{placement})$

\For{each node $v \in G_{HW}$}
    \If{$v$ is MVAU \textbf{and} $v \in M$}
        \State $v.\text{trojan\_enabled} \gets \text{true}$
        \State $v.\text{trojan\_config} \gets C$
        \State $v.\text{output\_routing} \gets \text{internal stream}$
    \EndIf
\EndFor

\State $G_{HLS} \gets \text{generate\_HLS}(G_{HW})$

\For{each MVAU\_HLS kernel $k \in G_{HLS}$ where $k.\text{trojan\_enabled}$}
    \State insert macro definitions from $k.\text{trojan\_config}$
    \State redirect output stream to internal buffer
    \State append \textsc{TrojanPostProcess} to compute loop
\EndFor

\State $R \gets \text{Vitis\_HLS\_synthesize}(G_{HLS})$
\State $B \gets \text{implement}(R, \theta)$

\State \Return $B$
\end{algorithmic}
\end{algorithm}

The implementation targets the final MVAU layer, which produces the logits used by the Label Select node for class prediction. The approach may apply to most CNN-based classification models compiled with FINN, as it depends only on a single MVAU layer. Modifying these logits enables controlled misclassification. The placement of the Trojan within the FINN inference pipeline is shown in Figure \ref{fig:trojan_placement}. The Input IODMA transfers the input through the unmodified layers until it reaches the final MVAU. When the trigger condition is met, the Trojan modifies the output logits before passing them to the Label Select node. For example, a logit-swap payload can exchange the scores of classes \textbf{3} and \textbf{7}, causing Label Select to select class \textbf{7} instead of the correct class \textbf{3}.
\begin{figure}[!htbp]
    \centering
    \includegraphics[width=\columnwidth]{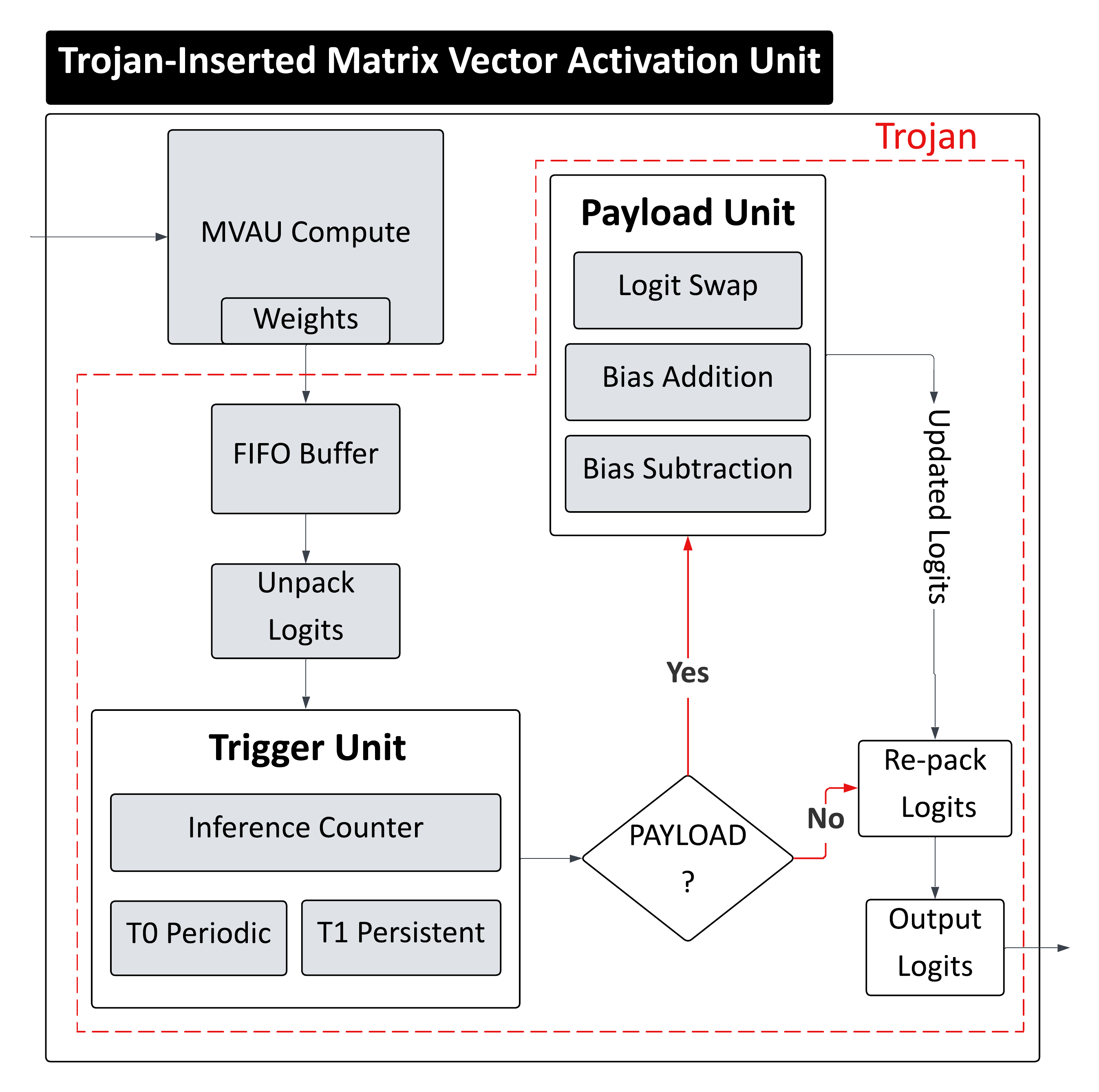}
    \caption{Block diagram of the Trojan-injected matrix vector activation unit (MVAU) layer in a FINN-based neural network accelerator.}
    \label{fig:mvau}
\end{figure}
\subsection{Trojan-inserted MVAU Layer}
A Trojan-inserted MVAU extends the standard FINN MVAU with additional trigger and payload logic within the same HLS-generated IP core. The original interface and computation are preserved, while an internal processing path conditionally modifies the output logits before transmission to downstream components. As shown in Fig. \ref{fig:mvau}, the modified MVAU consists of two functional components: the standard ``MVAU compute'' block and the Trojan logic, which contains the trigger and payload units.

The MVAU compute performs the standard matrix-vector multiplication, activation, and thresholding operations using the trained weights and thresholds. Unlike a standard MVAU, whose output is directly streamed downstream, the modified MVAU temporarily stores the output in an internal FIFO and reconstructs the complete logit vector for Trojan processing.
The trigger unit uses an inference counter and a latched flag to determine whether the payload is activated. In \textit{periodic} mode, the payload is activated every $N$-th inference and the counter is reset. In \textit{persistent} mode, activation occurs at inference $N$, after which the payload remains active for subsequent inferences. The trigger produces a Boolean activation signal for the payload unit.

Three payload modes are supported: \textit{bias addition}, \textit{logit swap}, and \textit{bias subtraction}. Bias addition increases the target-class logit by a specified value, whereas bias subtraction reduces it. Logit swap exchanges the logits of two selected classes to alter the classification decision. The trigger and payload logic are contained entirely within the MVAU IP and operate independently of downstream Top-K selection and runtime software control. Algorithm~\ref{alg:trojan_post_process} outlines the Trojan execution process.
\begin{algorithm}
\caption{: TrojanPostProcess (Trojan execution inside Trojan-inserted MVAU\_hls IP}
\label{alg:trojan_post_process}
\begin{algorithmic}[1]

\Require Stream $S$, trigger period $N$, trigger mode, payload mode, classes $c_t, c_s$, bias $b$

\State $L \gets \text{unpack}(S)$
\State $trigger\_count \gets 0$, $trojan\_latched \gets \text{false}$

\If{$trigger\_mode = \text{periodic}$}
    \State $fired \gets (trigger\_count = N-1)$
    \State update $trigger\_count$
\ElsIf{$trojan\_latched \lor trigger\_count = N-1$}
    \State $trojan\_latched \gets \text{true}$
    \State $fired \gets \text{true}$
    \State $trigger\_count \gets 0$
\Else
    \State update $trigger\_count$
\EndIf

\If{$fired$}
    \If{$payload\_mode = \text{bias\_addition}$}
        \State $L[c_t] \gets L[c_t] + b$
    \ElsIf{$payload\_mode = \text{logit\_swap}$}
        \State swap $L[c_t], L[c_s]$
    \ElsIf{$payload\_mode = \text{bias\_subtraction}$}
        \State $L[c_t] \gets \max(0, L[c_t] - b)$
    \EndIf
\EndIf

\State write $L$ to \texttt{out0\_V}

\end{algorithmic}
\end{algorithm}
\noindent

%% file: 4_Results.tex
\subsection{Experimental Setup}
An Ubuntu 24.04.4 LTS system with an Intel\textsuperscript{\textregistered} Core\texttrademark{} i7-13800H processor and 32 GB of RAM is used for compilation. Two DNN models are evaluated: a 4-layer feed-forward network for MNIST digit recognition and a 12-layer CNN for CIFAR-10 classification, as detailed in Tables \ref{tab:fcbnn_architecture} and \ref{tab:cnn_architecture} \cite{finn-examples}. For each model, we generate one baseline accelerator and six variants corresponding to the configurations in Table \ref{tab:trojan_configs}, using FINN (v0.10.1). All accelerators share identical hardware configurations, including zero DSP usage, identical folding parameters, HLS-based MVAU layers, and binary quantization across all layers.

\begin{table}[!htbp]
\centering
\caption{Layer-wise details of the MNIST digit recognition model}
\label{tab:fcbnn_architecture}
\begin{tabular}{|c|c|c|}
\hline
\textbf{Layer type} & \textbf{Layer} & \textbf{Data shape} \\
\hline
\multirow{2}{*}{Image} & Input-0 & {[}28, 28{]} \\
\cline{2-3}
 & Flatten-1 & {[}784{]} \\
\hline
\multirow{2}{*}{FC-1} & Linear-2 & {[}256{]} \\
\cline{2-3}
 & ReLU-3 & {[}256{]} \\
\hline
\multirow{2}{*}{FC-2} & Linear-4 & {[}256{]} \\
\cline{2-3}
 & ReLU-5 & {[}256{]} \\
\hline
FC-3 & Linear-6 & {[}10{]} \\
\hline
\end{tabular}
\end{table}

\begin{table}[!htbp]
\centering
\caption{Layer-wise details of the CNN-based CIFAR-10 Classification Model}
\label{tab:cnn_architecture}
\begin{tabular}{|c|c|c|}
\hline
\textbf{Layer type} & \textbf{Layer} & \textbf{Data shape} \\
\hline

\multirow{1}{*}{Image}
& Input-0 & {[}3, 32, 32{]} \\
\hline

\multirow{2}{*}{CONV-1}
& Conv-1 & {[}64, 30, 30{]} \\
& ReLU-2 & {[}64, 30, 30{]} \\
\hline

\multirow{2}{*}{CONV-2}
& Conv-3 & {[}64, 28, 28{]} \\
& ReLU-4 & {[}64, 28, 28{]} \\
\hline

POOL-1
& Pool-5 & {[}64, 14, 14{]} \\
\hline

\multirow{2}{*}{CONV-3}
& Conv-6 & {[}128, 12, 12{]} \\
& ReLU-7 & {[}128, 12, 12{]} \\
\hline

\multirow{2}{*}{CONV-4}
& Conv-8 & {[}128, 10, 10{]} \\
& ReLU-9 & {[}128, 10, 10{]} \\
\hline

POOL-2
& Pool-10 & {[}128, 5, 5{]} \\
\hline

\multirow{2}{*}{CONV-5}
& Conv-11 & {[}256, 3, 3{]} \\
& ReLU-12 & {[}256, 3, 3{]} \\
\hline

\multirow{2}{*}{CONV-6}
& Conv-13 & {[}256, 1, 1{]} \\
& ReLU-14 & {[}256, 1, 1{]} \\
\hline

\multirow{3}{*}{FC-1}
& Flatten-15 & {[}256{]} \\
& Linear-16 & {[}512{]} \\
& ReLU-17 & {[}512{]} \\
\hline

\multirow{2}{*}{FC-2}
& Linear-18 & {[}512{]} \\
& ReLU-19 & {[}512{]} \\
\hline

FC-3
& Linear-20 & {[}10{]} \\
\hline

\end{tabular}
\end{table}

\begin{table*}[htbp]
\centering
\caption{FINN-Tro Configurations Used to Demonstrate 
Verification Gaps in FINN}
\label{tab:trojan_configs}
\renewcommand{\arraystretch}{1.25}
\small
\begin{tabular}{|c|c|c|p{8.5cm}|}
\hline
\textbf{Run ID} & \textbf{Payload Type} & \textbf{Trigger Type} & 
\textbf{Description} \\ 
\hline
A1-T0 
& Bias Addition 
& Periodic 
& Every $N$-th inference, the payload forces the predicted class to 3. \\ 
\hline
A1-T1 
& Bias Addition 
& Persistent 
& The Trojan activates at the $N$-th inference and remains active thereafter, forcing all subsequent predictions to class 3. \\ 
\hline
A2-P1-T0
& Logit Swap 
& Periodic 
& At every $N$-th inference, the Trojan swaps the logits of two specified classes.\\ 
\hline
A2-P1-T1 
& Logit Swap 
& Persistent 
& The Trojan activates at the $N$-th inference and swaps two specified class logits thereafter. \\ 
\hline
A2-P2-T0 
& Bias Subtraction 
& Periodic 
& Every $N$-th inference, the Trojan subtracts a large bias from the class 3 logit. \\ 
\hline
A2-P2-T1 
& Bias Subtraction
& Persistent 
& The Trojan activates at the $N$-th inference and thereafter applies a large negative bias to the class 3 logit. \\ 
\hline
\end{tabular}
\end{table*}

\begin{table*}[ht]
\centering
\caption{Performance evaluation of FINN-Tro configurations on PYNQ-Z1}
\label{tab:perf_metrics}
\setlength{\tabcolsep}{5pt}
\renewcommand{\arraystretch}{1.15}
\small
\begin{tabular}{l ccc ccc}
\toprule
& \multicolumn{3}{c}{\textbf{MNIST Model}}
& \multicolumn{3}{c}{\textbf{CIFAR-10 Model}} \\
\cmidrule(lr){2-4} \cmidrule(lr){5-7}
\textbf{Run ID}
& \textbf{Acc. (\%)} & \textbf{Throughput (img/s)} & \textbf{Runtime (ms)}
& \textbf{Acc. (\%)} & \textbf{Throughput (img/s)} & \textbf{Runtime (ms)} \\
\midrule
\textbf{Baseline}
& \textbf{92.96} & \textbf{737,655} & \textbf{1.356}
& \textbf{84.19} & \textbf{1,995}   & \textbf{501.205} \\
A1-T0
& 76.20 & 737,007 & 1.357
& 69.80 & 1,988   & 502.940 \\
A1-T1
& 10.12 & 736,360 & 1.358
& 10.00 & 1,965   & 508.965 \\
A2-P1-T0
& 89.67 & 717,466 & 1.394
& 81.38 & 1,995   & 501.215 \\
A2-P1-T1
& 74.74 & 730,715 & 1.369
& 68.97 & 1,995   & 501.203 \\
A2-P2-T0
& 91.43 & 711,864 & 1.405
& 83.29 & 1,995   & 501.200 \\
A2-P2-T1
& 84.33 & 720,919 & 1.387
& 78.84 & 1,995   & 501.214 \\
\bottomrule
\end{tabular}
\end{table*}

The resulting accelerators are deployed on the Xilinx PYNQ-Z1 Evaluation Board, which contains an AMD Zynq xc7z020-1clg400 FPGA and a dual-core ARM Cortex-A9 processor. Performance is evaluated at a fixed clock frequency of 100 MHz and a batch size of 1000 images using accuracy, throughput, and runtime as metrics.

Table \ref{tab:trojan_configs} summarizes the six FINN-Tro configurations obtained by combining three payloads—Bias Addition, Logit Swap, and Bias Subtraction—with two trigger modes: \textit{Periodic}, which activates every $N$-th inference, and \textit{Persistent}, which activates at the $N$-th inference and remains active thereafter. For reproducibility, (N) and (b) are fixed at 5 and 255, respectively. As compile-time constants (Algorithm~\ref{alg:trojan_insertion}, line 6), they have negligible resource overhead; an adversary could instead use a larger N or smaller b to remain within the accuracy tolerance and time budget. These configurations demonstrate that the FINN verification gap can be exploited through different Trojan behaviors.

\subsection{Performance Evaluation}
Table \ref{tab:perf_metrics} compares the baseline and six FINN-Tro configurations on MNIST and CIFAR-10. The baseline achieves 92.96\% accuracy with 737,655 images/s throughput and 1.356 ms runtime for MNIST, while the CIFAR-10 model achieves 84.19\% accuracy, 1,995 images/s throughput, and 501.205 ms runtime. Across all Trojan configurations, throughput and runtime remain close to the baseline, indicating low latency overhead. In contrast, accuracy degradation depends strongly on the Trojan configuration. The persistent Bias Addition attack (A1-T1) produces the largest degradation, reducing accuracy to 10.12\% on MNIST and 10.00\% on CIFAR-10. Periodic configurations produce more moderate degradation because the payload is activated only every $N$-th inference. These results demonstrate that FINN-Tro can induce controlled misclassification while preserving the accelerator's normal performance characteristics.

\subsection{FPGA Resource Utilization}
To quantify the overhead of each attack variant, post-implementation resource utilization and on-chip power consumption are evaluated, as shown in Table \ref{tab:resource_consumption}. The MNIST baseline uses 25.83\% of available LUTs, 4.08\% LUTRAM, 18.80\% FFs, and 10.36\% BRAM, with 1.642 W of on-chip power. The larger CIFAR-10 CNN uses 47.87\% LUTs, 11.70\% LUTRAM, 29.70\% FFs, and 71.07\% BRAM, with 2.085 W of power.

Trojan insertion introduces only minor resource changes. Compared with the baseline, LUT utilization increases by at most 6.71\% for MNIST and 2.50\% for CIFAR-10, while FF usage increases by at most 7.49\% and 3.98\%, respectively. On-chip power remains within 9 mW of the respective baselines. LUTRAM increases by at most 4.51\% for MNIST and 0.05\% for CIFAR-10, while BRAM increases by at most 6.90\% and 1.01\%, respectively. Notably, the relative overhead decreases with increasing network depth.

Overall, FINN-Tro introduces very small hardware and power overhead while maintaining performance comparable to the baseline accelerators, making the Trojan variants difficult to distinguish through conventional resource and power profiling.

\begin{table}[!htbp]
\centering
\caption{FPGA resource utilization and on-chip power across FINN-Tro attack configurations}
\label{tab:resource_consumption}
\setlength{\tabcolsep}{5pt}
\renewcommand{\arraystretch}{1.12}
\small
\resizebox{\columnwidth}{!}{%
\begin{tabular}{l c c c c c}
\toprule
\textbf{Run ID} & \textbf{On-chip Power (W)} & \textbf{LUT} & \textbf{LUTRAM} & \textbf{FF} & \textbf{BRAM} \\
\midrule
\multicolumn{6}{l}{\textit{MNIST Model}} \\
\midrule
\textbf{Baseline} &
\textbf{1.642} &
\textbf{13743} &
\textbf{709} &
\textbf{20012} &
\textbf{14.5} \\
A1-T0 &
$1.636$ & $14635$ & $741$ & $21476$ & $14.5$ \\
A1-T1 &
$1.634$ & $14639$ & $741$ & $21478$ & $14.5$ \\
A2-P1-T0 &
$1.640$ & $14596$ & $709$ & $21444$ & $15.5$ \\
A2-P1-T1 &
$1.639$ & $14593$ & $709$ & $21446$ & $15.5$ \\
A2-P2-T0 &
$1.633$ & $14658$ & $741$ & $21508$ & $14.5$ \\
A2-P2-T1 &
$1.634$ & $14665$ & $741$ & $21510$ & $14.5$ \\
\midrule
\multicolumn{6}{l}{\textit{CIFAR-10 Model}} \\
\midrule
\textbf{Baseline} &
\textbf{2.085} &
\textbf{25465} &
\textbf{2035} &
\textbf{31599} &
\textbf{99.5} \\
A1-T0 &
$2.088$ & $25676$ & $2036$ & $32286$ & $100.5$ \\
A1-T1 &
$2.076$ & $26100$ & $2022$ & $32855$ & $99.5$ \\
A2-P1-T0 &
$2.081$ & $25689$ & $2035$ & $32285$ & $100.5$ \\
A2-P1-T1 &
$2.074$ & $25713$ & $2036$ & $32287$ & $100.5$ \\
A2-P2-T0 &
$2.084$ & $25701$ & $2036$ & $32319$ & $100.5$ \\
A2-P2-T1 &
$2.085$ & $25711$ & $2036$ & $32321$ & $100.5$ \\
\bottomrule
\end{tabular}%
}
\end{table}

%% file: 5_Conclusion.tex
This paper presented FINN-Tro, a hardware Trojan insertion methodology that exploits a verification gap in the FINN compilation pipeline to embed malicious logic without modifying the original quantized model or trained weights. The Trojan is inserted into the final MVAU layer during HLS code generation and combines two counter-based trigger modes, \textit{Periodic} and \textit{Persistent}, with three payloads: \textit{Bias Addition}, \textit{Logit Swap}, and \textit{Bias Subtraction}, resulting in six different Trojan configurations.
Our results show that the Trojan remained undetected throughout the design flow and was successfully activated on a PYNQ-Z1 board, causing the intended accuracy degradation while maintaining near-baseline throughput and runtime with only modest hardware overhead, thus motivating stronger design-time checks in current verification of automated accelerator-generation flows.

Future work will extend FINN-Tro to deeper network architectures, investigate input-conditioned trigger mechanisms, and evaluate verification checkers as potential countermeasures.